# Nodes having a major influence to break cooperation define a novel centrality measure: game centrality

Gabor I. Simko[1,2,*], Peter Csermely[1,**]

**1** Department of Medical Chemistry, Semmelweis University, Budapest, Hungary, **2** Vanderbilt University, Nashville, TN USA

**Abstract**

**Cooperation played a significant role in the self-organization and evolution of living organisms. Both network topology and the initial position of cooperators heavily affect the cooperation of social dilemma games. We developed a novel simulation program package, called 'NetworGame', which is able to simulate any type of social dilemma games on any model, or real world networks with any assignment of initial cooperation or defection strategies to network nodes. The ability of initially defecting single nodes to break overall cooperation was called as 'game centrality'. The efficiency of this measure was verified on well-known social networks, and was extended to 'protein games', i.e. the simulation of cooperation between proteins, or their amino acids. Hubs and in particular, party hubs of yeast protein-protein interaction networks had a large influence to convert the cooperation of other nodes to defection. Simulations on methionyl-tRNA synthetase protein structure network indicated an increased influence of nodes belonging to intra-protein signaling pathways on breaking cooperation. The efficiency of single, initially defecting nodes to convert the cooperation of other nodes to defection in social dilemma games may be an important measure to predict the importance of nodes in the integration and regulation of complex systems. Game centrality may help to design more efficient interventions to cellular networks (in forms of drugs), to ecosystems and social networks.**

---

[*] The work published in this paper was done while the author worked at the Budapest University of Technology and Economics, Budapest, Hungary.
[**] Corresponding author; Department of Medical Chemistry, Semmelweis University, P.O.Box 260, H-1444 Budapest 8, Hungary; Tel: +36-1-459-1500/60130; E-mail: csermely.peter@med.semmelweis-univ.hu

## Introduction

Cooperation is necessary for the emergence of complex, hierarchical biological organisms. Prisoner's dilemma and hawk-dove games are social dilemma games played by agent-pairs having two strategies: cooperation or defection. These games are excellent models to elucidate the appearance of cooperation in situations, when agents generally prefer defection. Such situations are prevalent in evolutionary biology, where personal interests often confront with collective interests [1,2]. Recent reports defined 'signaling games' [3-6] or 'protein games' [7-9]. In protein games the applicability of social games was extended to protein-protein interaction networks, where cooperation level of the whole system helped to assess the overall integration of network functions. Here we extend these ideas further, and show the applicability of the spatial game concept to protein structure networks (also called residue interaction networks) having amino acids as their nodes.

In spatial games only those agents are playing with each other, who are neighbors in the underlying contact network. The level of cooperation is very sensitive to the topology of the agent network in a wide variety of social dilemma type games. Cooperation levels in square lattice, small world and scale-free network topologies have been widely assessed [10-17]. However, only a few studies were performed on real world networks [18,19] and even less studies examined the effects of pre-set starting cooperative and defective strategies on the development of cooperation at the network level [20,21]. Although, there are a number of spatial game-related programs [22-25], but none of them is able to simulate social dilemma games on real world networks with pre-set initial strategies.

Following our conference report on an initial version of the program [26], here we introduce the improved version of a freely available simulation tool, called NetworGame able to simulate two-player, pairwise interacting social dilemma games on real world networks with any assignment of initial cooperation or defection strategies to network nodes. A novel dynamic centrality, called game centrality, is also defined, which measures the ability of individually defecting nodes to convert others to their strategy. We show the applicability of game centrality on social, protein-protein interaction and protein structure networks, and highlight the importance of hubs in the maintenance of cooperation in complex biological systems.

## Results

**NetworGame program for simulation of spatial games on any real-world networks**
Our novel NetworGame 2.0 program package is a cross-platform, generic tool to simulate repeated spatial games. This simulation program includes i.) options for pay-off matrices of any symmetric normal form games (with 2 strategies); ii.) well-known, replicator-type strategy update rules (best takes over, Fermi-rule and proportional update [13]), as well as the option for additional, user-defined strategy update rules in a 'plugin'-type format; iii.) synchronous, and semi-synchronous updating [27]; iv.) and the option for the inclusion of any real world networks in a Pajek [28] format (for the description of the simulation steps see **Text S1**). The program allows setting the starting strategy of any nodes according to the wish of the experimenter, and introduces the novel use of edge weights by considering edge weights as probabilities of the game played between the corresponding nodes. A brief description of the NetworGame 2.0 software is given in the Methods section, a more detailed description containing the pseudocode of the algorithm used is found in the Electronic Supplementary Material (**Text S1**). The program with a User Guide is freely available from our website: www.linkgroup.hu/NetworGame.php.

**Definition of game centrality as a relative importance of nodes to maintain cooperation**
Keeping in mind the relative scarcity of dynamic centrality measures [29-33] we defined a novel dynamic centrality, called game centrality, as follows. Initially let all nodes but node $i$ cooperate, while set the initial strategy of node $i$ to defect. Under these initial conditions the game centrality of a node $i$ ($GC_i$) is equal to the proportion of defectors averaged over the last 50 simulation steps.

In the determination of the simulation step range of the averaging process we considered two opposing effects. 1.) Averaging over a large number of steps resulted in more accurate results. 2.) However, averaging more game rounds also hindered convergence of the mean game centrality value, which determined the number of total rounds of the repeated game as described in the next paragraph. As we will discuss in the following paragraphs, game centrality (GC) is a relative measure useful for the comparison of the nodes in the same network. Therefore, GC may also be calculated and compared using more or less steps of average than 50. However, the selection of the last 50 game rounds as the 'average-window' avoided both potential pitfalls.

Game centrality values depend on the number of time steps (game rounds) and the number of parallel simulations. We may minimize GC-variability by measuring the proportion of defectors, where already no large fluctuations can be seen. This condition can be reached most of the times by using prisoner's dilemma game on real world networks (but can not be reached using e.g. the evolutionary prisoner's dilemma game on a square lattice [11]). Ensuring this condition, in the simulations of this paper the repeated rounds of serial games were concluded, when the mean value of GC changed less than 0.01 (called as GC-drift threshold) in the last 50 rounds. Small fluctuations caused by the stochasticity of some strategy update rules (e.g. replicator dynamics) and by the occasional stochasticity of edge weights (i.e. using the edge weights as update probabilities) may be minimized by averaging several simulations. The number of simulations was chosen to ensure that the mean error for the proportion of defectors was below 0.01 in the final step (called as GC-fluctuation threshold). As noted earlier, GC is used for the comparison of nodes. Therefore, both the GC-drift and GC-fluctuation thresholds may be set different than 0.01. However, selection of the threshold value 0.01 resulted in enough rounds of game simulations to surpass the initial transitional phase, often observed in repeated games, and resulted in relatively stable game centrality values. We note that the NetworGame simulation program can also be set to run simulations having a user-defined, pre-set number of game rounds.

It is important to note that numerical values of GC also depend on the payoff conditions and payoff parameters of the actual game model, on the applied strategy update rule, on the synchronicity of the update, as well as on the network structure. These features, together with those mentioned earlier, make GC a relative measure for the comparison of nodes in the same network using the same game conditions. Additionally, we found that the ranking of node GC-s largely correlated for different strategy update rules and temptation values for the relatively large yeast protein-protein interaction network studied (having 2,444 nodes and 6,271 edges; see data later). In this sense GC may also be used for a rough comparison of node importance to break cooperation using different game conditions.

In the current simulations we chose the canonical prisoner's dilemma game with the payoff parameters R=3, T=6, S=0, P=1 (except for Michael's strike network, where the temptation value was less: T=3.1), since selection of the maximal temptation (T) value ensured the largest sensitivity of the initially almost fully cooperating network to the defection of a single node. Simulations used the widely applied best takes over strategy update rule with synchronous update.

**Game centrality identifies influential members of Zachary's karate club network**
Wayne Zachary [34] recorded the strength of contacts between members of a university karate club between 1970 and 1972. Meanwhile, the club had a dispute between the club president and the chief karate instructor, leading to a fission resulting in two separate clubs, which made this social network a gold standard for network modularization studies. To determine the most influential members of the karate club using the game centrality measure (GC) defined above, we simulated a prisoner's dilemma game on the Zachary karate club network with the initial cooperation of every node except the examined, defective club member. Nodes #3 and #33, as well as #1, #2 and #34 had the top GC values having equal GC values within the first and second group of nodes and decreasing GC in the second group as compared to that of the first group. Node #1 corresponded to the instructor, while node #34 represented the club president. The large efficiency of these two and the other 3 nodes to break cooperation is related to the fact that the 5 nodes listed above had the five highest degrees in the network, and they were also found among the seven nodes having the largest betweenness centrality values (node ID-s in the order of decreasing centralities: #1, #34, #33, #3, #32, #9, #2).

**Game centrality measures identify influential member-pairs of Michael's strike network**
As a next step, we were interested, whether the game centrality of edges (i.e. the average proportion of defectors in the last 50 rounds, where not only a single node, but two neighboring nodes are both initial defectors) is also giving meaningful results. Michael's strike network [35] was an excellent example to test this measure. Judd H. Michael described a strike in a forest product manufacturing factory. The factory had a new management, who wanted to change the compensation package of the workers. The two union negotiators (Sam and Wendle) were responsible for explaining the changes, but they failed to do so, and a strike broke out. The company hired Judd H. Michael to make a sociogram, which showed that there were three worker groups: younger, English-speaking, older, English-speaking, and younger, Spanish-speaking workers. Sam and Wendle formed a linked pair of nodes. Judd H. Michael advised to contact another linked pair of nodes, Bob and Norm – who were at the overlap of the three communities of the factory sociogram (see **Figure S1 of Text S1**) –, and to convince them about the changes. By following this strategy, the management solved the problem soon, and the strike ended.

In the social dilemma game simulations of the situation we considered strikers as cooperators and strike-breakers as defectors in canonical prisoner's dilemma games having the payoff parameters of R=3, T=3.1, S=0, P=1. Initially everybody was cooperating, but the two linked workers chosen to explain the changes to the others. In repeated simulations the influence of different negotiator-pairs was compared. Simulation of the choice of Bob and Norm showed that they could convince everybody to stop the strike in 100% of the simulations. Simulation of the choice of Sam and Wendle led to the poor result of convincing the others to stop the strike in 8% of the simulations with the same settings, which corresponds well with outcome of the real-world events [35] (we got the same results for the weak prisoner's dilemma game;

data not shown). Our results show, that not only node game centrality, but also edge game centrality is giving a similar outcome than those happened in real-world situations.

**Game centrality correlates with former centrality measures and reveals novel centers of influence in protein-protein interaction networks**

As two proteins approach each other, they signal their status to the other *via* the hydrogen-bonded network of water molecules. Binding is achieved by a complex set of consecutive conformational adjustments. These concerted, conditional steps were called as a 'protein dance', and can be perceived as rounds of a repeated game [6-9]. Here we used the canonical prisoner's dilemma game with a maximal temptation value (T=6), since these parameters represent the most stringent conditions of cooperation among the most commonly used social dilemma games.

First we examined the effect of defection of party and date hubs on the cooperation of the high-fidelity yeast (*Saccharomyces cerevisiae*) protein-protein interaction network (interactome) of Ekman *et al.* [36] using the canonical prisoner's dilemma game. Party hubs are hubs, which do not change their neighborhood structure, and are often situated in the middle of network modules. On the contrary, date hubs change their neighbors frequently, and often connect various modules of the interactome. The best distinction between party hubs and date hubs has been a subject of recent discussions [37-43]. Prisoner's dilemma game simulations confirmed the differences between the two types of hubs. In the simulations initially we let all the 2,444 nodes cooperate except for 30 nodes, which defected. We compared the average game centrality values over 2000 simulations by random sampling 30 defecting nodes from the 63 consensus party hubs (compiled as in [43], see **Table S1 of Text S1**), or from the 145 consensus date hubs (compiled as in [43], see **Table S2 of Text S1**), or from all the 2,444 nodes, respectively. Party hubs had the largest game centrality, while date hubs and randomly selected node sets had smaller and smaller game centrality values, i.e. they distorted less and less the initial cooperation (**Table 1**). Using the chi-square test we found that the distribution of game centralities were significantly different ($\chi^2 > 400$) for the different test cases.

Next, we determined the correlation between degrees, betweenness centralities and GC-s in prisoner's dilemma games of the 2,444 yeast proteins of the high fidelity yeast interactome [36]. Since both degree and betweenness centrality had a large number of tied values, we used the Goodman-Kruskal gamma test to test the association and significance of the results. Using canonical prisoner's dilemma game we found that GC has a good correlation with both degrees and betweenness centralities (**Table 2**). We were also curious, which of the 3 measures of degree, betweenness centrality or game centrality predicts better the phenotypic potential of yeast protein describing their 'buffering capacity' against evolutionary changes, i.e. the contribution of an individual yeast protein to the overall robustness of yeast cells [44]. GCs were found to be significantly ($p<0.062$) better predictors of genetic buffering of evolutionary changes than either degrees or betwenness centralities (**Table 2**).

The functional analysis of the 171 proteins of the high-fidelity yeast interactome [36] causing the final cooperation level to fall from the starting value of close to 1.0 to less than 0.9 showed the overrepresentation ($p<0.001$) of nucleus-related functions (35%). The second and third most overrepresented classes were signaling- and transport-related functions (33% and 32%, respectively; **Figure 1**). These results were in compliance with the central position of the nucleus in the structure, organization and dynamics of the cell. Similarly, transport and signaling are central in both internal and external cellular communication.

In conclusion, game centrality observed in prisoner's dilemma games of nodes in a yeast protein-protein interaction network (i.e. in a 'protein game') offered a novel characterization of the importance of proteins in complex cellular functions highlighting the importance of intra-modular party hubs to maintain cooperation.

As we said before, game centrality is a comparative measure within a network having set a game type, strategy update rule, temptation value and synchronicity. To assess the consistency of game centrality ranking, we ran multiple experiments with different temptation values (T=3.1, 4, 5 and 6) and/or the strategy update rules (best-takes-over and Fermi-rule) for the yeast interactome network. We evaluated the pair-wise correlation (Goodman-Kruskal gamma) between the game centralities for the case of small (3.1 and 4.0) and large temptation values (5.0 and 6.0), while applying the best-takes-over and the Fermi-rule strategy update rules. The smallest pair-wise correlation was 0.72 for the small temptation values and 0.70 for the large temptation values. These results indicate that while the game settings do have effects on the individual game-centrality values, GC may also be used for a rough comparison of node importance to break cooperation using different game conditions.

**Game centrality identifies functionally important segments of protein structures**
Next, we extended the use of the game centrality to another important biological network, the protein structure network, where the nodes are amino acids, and the edges between them represent chemical bonds [6,45,46]. We analyzed the protein structure network of the *Escherichia coli* methionyl-tRNA synthetase protein, for which an elegant study [46] showed the existence of several alternative intra-protein signal transduction pathways. These signaling paths span a large distance between the active centre and the anticodon binding region of this enzyme transmitting the allosteric conformational changes induced by substrate binding. The methionyl-tRNA synthetase protein has two major domains, the catalytic domain (responsible for the activation of methionine) and the tRNA anticodon-binding domain. These two domains are connected by the connecting peptide (CP) domain. The catalytic domain can be subdivided to three sub-domains, having two Rossmann-folds and a stem contact fold [46].

First, we compared the game centrality (GC) of the amino acids in the two major domains and their connecting peptide. The average GC values of both major domains were higher (both before and after substrate binding) than that of the connecting peptide domain (**Table 3**, **Figure 2**). In agreement with their central role in protein function, the average GC of intra-protein signaling amino acids as defined by Ghosh and Vishveshwara [46] was especially high, if compared to GC-s of the rest of the amino acids (**Table 3**). Substrate binding induced a decrease of GC of most domains, which is in agreement with the development of a more compact structure, where amino acids may indeed have a lower individual influence on domain-level processes. GC differences between the open and closed conformations reflected that substrate binding affected most the tRNA anticodon domain and the core of the catalytic domain (the first Rossmann-fold domain; **Table 3**, **Figure 2**), which is again in agreement with the high increase of compactness around the tRNA and substrate binding pockets upon substrate binding. These findings are also in agreement with the prominent influence of binding sites on the cooperating network of amino acids revealed by molecular dynamics simulations [47].

In conclusion, game centrality values observed in prisoner's dilemma games of individual amino acids of protein structure networks highlighted the importance of core protein domains, especially the tRNA anticodon binding domain, the active centre and intra-protein signaling amino acids in the maintenance of cooperation of this complex system.

**Discussion**

In this paper a novel program package, called NetworGame was introduced to simulate various social dilemma type games with a high flexibility for payoff conditions, initial parameters, strategies, strategy update rules and update conditions. We defined a novel dynamic centrality measure called game centrality (GC), and showed that it correlates with previous centrality measures, such as degree, or betwenness centralities. Moreover, GC also predicts novel influential nodes and network segments, which have a functional relevance in several social and biological real world networks.

Although several game simulation tools have been described in the literature, such as GamePlan [22], Gambit [23], Dynamo [24] and VirtualLabs/EvoLudo [25], the NetworGame program package is unique in the sense, that it is able to accept any real-world networks as an input with the complexity of their weighted edges. Moreover, the program can handle individual initial strategies of all networked agents, as well as individual strategy update rules. The current NetworGame 2.0 version has automated, statistics-based simulation length and simulation count options, which were not present in its preliminary NetworGame 1.0 version mentioned in our former conference report [26]. We note that NetworGame 2.0 can also be used for simulations having a user-defined, pre-set number of game rounds.

While game centrality correlated with previous centrality measures, such as degree and betweenness centrality, it also predicted novel nodes and centers of large influence, which had functional relevance both in social and biological real-world networks. Identification of functionally and dynamically important network nodes and segments is not an easy field. i.) The identification of nodes with large and dynamic influence has been notoriously difficult [29-33, 48-50]. ii.) The precise discrimination between hubs with different dynamic parameters became a subject of recent discussions [36-43]. iii.) The contribution of individual proteins to the overall robustness of the cell against evolutionary changes is a largely unresolved question [44, 51, 52]. iv.) Though enzyme active sites and protein binding hot-spots have been identified using network metrics, such as betweenness centrality, amino acids of intra-protein signaling are not readily distinguishable [53-55]. Game centrality provides a novel and promising aspect to compare the influence of network nodes and segments using the cooperation-related, complex dynamic background of social dilemma games.

Game centrality values depend on the game model and its payoff conditions, on the applied strategy update rule, on the synchronicity of the update, on the network structure as well as on the number of time steps and total number of simulations. This makes game centrality a relative measure, and requires an especially large level of cautiousness in its use, when the level of final cooperation fluctuates (e.g. in games on square lattice [11], rock-scissors-paper games [56], etc.). However, the case studies we made suggest that game centrality values of both individual nodes and edges (pairs of neighboring nodes) can be useful for the comparison of influence in the same network when using the canonical prisoner's dilemma game on various networks. Moreover, game centrality may be used for the comparison of different states of the same network, which differ e.g. only in their weight structure, like that of the yeast protein-protein interaction network before and after stress [57].

In this paper we calculated game centrality for initially defecting single nodes or edges (pairs of neighboring nodes) of the network. However, it is important to note that a similar centrality measure may also be calculated for the inverse situation, where a pair of linked nodes is the only initial cooperator in an otherwise defecting network. Moreover, similar centrality measures may also be defined for larger segments of nodes, such as for triangles, motifs, *k*-cliques, *k*-clans, *k*-clubs, *k*-components, *k*-plexes, lambda-sets, network skeletons, rich clubs, network cores, or network communities [43] using the NetworGame program.

From a different point of view, game centrality indirectly measures the capability of a node to alter network reciprocity. A highly influential node with large game centrality tends to break the cooperative islands, the sources of network reciprocity in the network. In this paper we do not discuss network reciprocity in detail, but we refer the interested reader to references [58] and [59].

Currently NetworGame only supports pairwise interactions between the players. It is an interesting future work to include group interactions in NetworGame, since it is known [60] that group interaction can lead to behaviors that cannot be attributed to the sum of pairwise interactions.

In conclusion, our NetworGame program package proved to be a useful tool for the analysis of repeated spatial games on real-world networks, and enabled the definition of a novel game-related centrality measure called game centrality (GC). Game centrality correlated with previous centrality measures, such as degree and betweenness centrality. Moreover, GC also identified novel, functionally important nodes and network segments in both social and biological networks. Our work opens the ground for a number of further studies on the dependence of game centrality of various parameters of social dilemma games as well as a wide variety of real world networks including neuronal networks [61]. Game centrality may become a useful measure to identify key network nodes of various processes of network dynamics, such as conformational changes, allosteric and cellular signaling, cell differentiation, cell reprogramming and malignant transformation. Game centrality may help to design more efficient interventions to cellular networks (in forms of drugs), to ecosystems or to social networks.

## Methods

**Description of the real world networks**
*Zachary's karate club network.* The weighted and undirected social network of the karate club at a US university contained 34 nodes and 78 edges described by Wayne Zachary in 1977 [34]. *Michael's strike network.* The social network of a forest product manufacturing factory contained 24 nodes and 38 edges as described by Judd H. Michael in 1997 [35]. *Yeast protein-protein interaction network.* The giant component of the un-weighted and undirected high-fidelity yeast protein-protein interaction network [36] contained 2,444 nodes and 6,271 edges, covering approximately half of the yeast genome and containing the most reliable ~3% of the expected number of total edges. *E. coli methionyl-tRNA synthase protein structure network.* We have constructed the structural network from the 3D image of the starting (substrate-free) and the equilibrated (substrate-bound) state of the molecular simulation of the *E. coli* methionyl-tRNA synthetase/tRNA/MetAMP complex [46] by converting the Cartesian coordinates of the 3D image to distances of amino acid pairs, and keeping all non-covalently bonded contacts within a distance of 0.4 nm. The final weighted network was created by removing self-loops, and calculating the inverse of the average distance between amino acid residues as edge weights. The protein structure network contained 547 nodes, since the first 3 N-terminal amino acids were not participating in the network. The protein structure contained 2,164 edges in the substrate-free network, and 2,153 edges in the substrate-bound network.

**Brief description of the NetworGame program package**

The NetworGame software is a program that takes a configuration specification describing the social dilemma game rules, the model or real world network and other settings, executes the simulations accordingly, and stores the simulation results. The simulation engine of the NetworGame program is a highly optimized software component for running the actual simulation. The 2.0 version of the NetworGame program is an updated version of the NetworGame 1.0 version published in a preliminary conference report [26]. A more detailed description of the NetworGame program version 2.0 containing the pseudo codes of its algorithms is found in the Supporting Information (**Text S1**). Both versions of the NetworGame program, as well as their User Guides are freely downloadable from our website: www.linkgroup.hu/NetworGame.php.

**Parameters of simulations and calculation of game centrality (GC)**

For the analysis of real-world networks we chose the canonical prisoner's dilemma game with the payoff parameters R=3, T=6, S=0, P=1 (except for Michael's strike network, where we used T=3), since this selection with a maximal temptation (T) value ensured the largest sensitivity of the initially almost fully cooperating network to the defection of a single node. In our simulations edge weights were not used. All simulations used the widely applied best takes over strategy update rule with a synchronous update. Simulations of repeated games were halted, when the mean value of cooperation changed less than 0.01 in the last 50 rounds. The number of parallel simulations was chosen to ensure that the mean error for the cooperation level was below 0.01 in the final step. These conditions allowed enough simulations to get a statistically meaningful mean estimate, and made the number of simulation steps large enough to surpass the initial transitional phase often observed in simulations of repeated social dilemma games. Game centrality (GC) of node *i* was calculated as the proportion of defectors averaged over the last 50 simulation steps, when initially node *i* was set to defect, while all other nodes were set to cooperate.

# Supporting information

**Text S1** This supporting information (Text S1) contains a supplementary figure, 2 supplementary tables, a detailed description of the NetworGame algorithm for the simulation of spatial social dilemma games with pseudocodes, as well as 11 supplementary references.

# Acknowledgements

The authors would like to thank Saraswathi Vishveshwara (Molecular Biophysics Unit, Indian Institute of Science, Bangalore) and Amit Ghosh (Lawrence Berkeley National Laboratory, Berkeley CA, USA) for the 3D data of the molecular simulation of the *E. coli* methionyl-tRNA synthetase/tRNA/MetAMP complex and Eszter Hazai (VirtuaDrug Co., Budapest, Hungary) for the protein structure network data. We thank Csaba Böde (Morgan Stanley, Budapest, Hungary) and István A. Kovács (Wigner Research Centre for Physics, Hungarian Academy of Sciences, Budapest, Hungary) for their help in the statistical calculations. We thank members of the LINK-group (www.linkgroup.hu) especially Miklós

## Tables

**Table 1.** Game centrality of party hubs, date hubs and randomly selected nodes of a high-fidelity yeast protein-protein interaction network.

|  | Consensus party hubs[b] | Consensus date hubs[b] | Randomly selected nodes[b] |
|---|---|---|---|
| Average Game Centrality (GC) of node sets[a] | 0.789±0.001[c] | 0.720±0.003 | 0.658±0.005 |

[a]Prisoner's dilemma game was simulated using the high-fidelity yeast interactome of Ekman *et al.* [36], and game centrality measures were calculated as described in Methods.
[b]Initially all 2,444 nodes were cooperating except for 30 defecting nodes, which were randomly sampled 2000 times from 63 consensus party hubs (compiled as in [43], see Table S1 of Text S1), from 145 consensus date hubs (compiled as in [43], see Table S2 of Text S1), as well as from all the 2,444 nodes in the network.
[c]Data represent sample means ± standard error. The distributions of the game centrality values were significantly different according to the chi-square test ($\chi^2 > 400$).

**Table 2.** Correlations of game centrality (GC) with degree, betweenness centrality and phenotypic potential of proteins in a high fidelity yeast interactome.

| Correlation values[a] | Degree | Betweenness Centrality | Game Centrality (GC) |
|---|---|---|---|
| Degree | --- | 0.81±0.02[b] (p<0.001) | 0.61±0.04 (p<0.001) |
| Betweenness Centrality | --- | --- | 0.62±0.04 (p<0.001) |
| Phenotypic potential | 0.09±0.04 (p<0.022) | 0.07±0.04 (p<0.083) | 0.13±0.05[c] (p<0.007) |

[a]Simulations of the prisoner's dilemma game were performed as described in Methods using the parameter set of (R=3, T=6, S=0, P=1). Correlation values between degree, betweenness centrality, GC in prisoner's dilemma game, as well as phenotypic potential [44] were calculated for the 2,444 proteins of the high fidelity yeast interactome of Ekman *et al.* [36].

[b]Data represent Goodman-Kruskal's gamma values ± standard errors. Significance levels in parentheses were also calculated using Goodman-Kruskal's gamma test (the null hypothesis being that the correlation is different from zero).

[c]Using the R-package correlation test (http://personality-project.org/r/html/r.test.html, [62]) the correlation between phenotypic potential and game centrality was significantly larger than the correlation between phenotypic potential and degree, or the correlation between phenotypic potential and betweenness centrality.

**Table 3.** Average game centrality (GC) values for *E. coli* methionyl-tRNA synthetase amino acids.

|  | Average Game Centrality (GC) of substrate-free protein[a] | Average Game Centrality (GC) of substrate-bound protein | Game Centrality (GC) decrease |
|---|---|---|---|
| Catalytic domain[b] | 0.69 | 0.60 | 0.09 |
| • *Rossmann-fold-1(active centre)* | *0.73* | *0.56* | *0.17* |
| • *Rossmann-fold-2* | *0.68* | *0.62* | *0.06* |
| • *Stem contact fold (KSMKS)* | *0.65* | *0.61* | *0.04* |
| Connecting peptide (CP) domain | 0.50 | 0.36 | 0.14 |
| tRNA anticodon-binding domain | 0.63 | 0.41 | 0.22 |
| Signaling amino acids [39] | 0.79 | 0.73 | 0.06 |
| Complete Met-tRNA-synthetase | 0.62 | 0.47 | 0.15 |

[a]Protein structure network of *E coli* methionyl-tRNA-synthetase was constructed, Prisoner's dilemma game was simulated, and game centrality measures were calculated as described in Methods.

[b]Domains from top to bottom: the catalytic domain including the Rossmann-fold-1 (catalytic function), Rossmann-fold-2 and stem contact fold (KMSKS) sub-domains; the connecting peptide (CP) domain; the anticodon binding, carboxy-terminal domain, 43 signaling amino acids involved in the transmission of conformational change as shown by Ghosh and Vishveshwara [46], whole methionyl-tRNA synthetase.

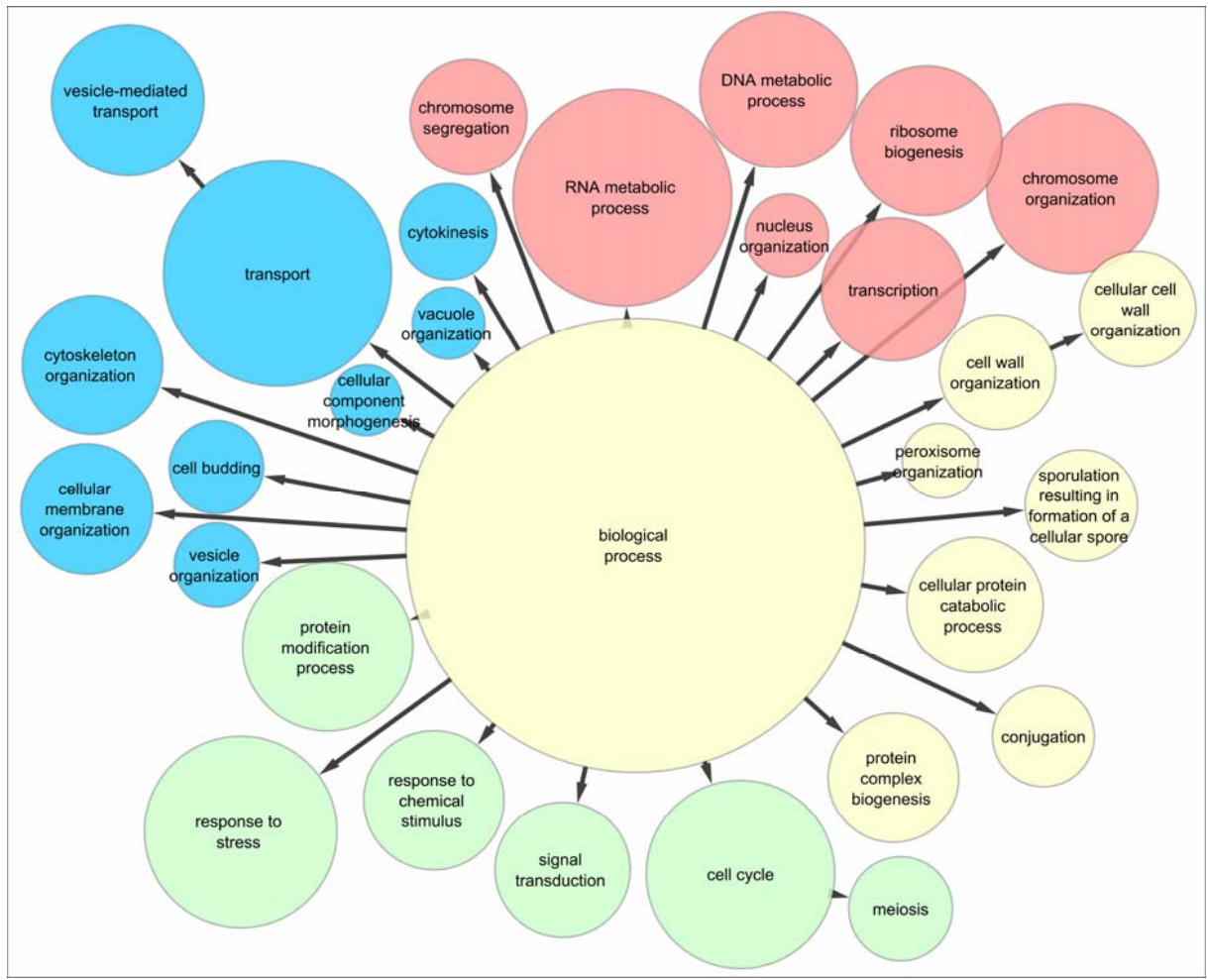

**Figure 1. Functional analysis of yeast proteins having the largest game centralities.**
Prisoner's dilemma game was simulated on a high-fidelity yeast interactome [36], and game centrality measures were calculated as described in Methods. 171 proteins out of the 2,444 nodes of the high-fidelity yeast interactome were selected by selecting nodes, which diminished the cooperation level from ~1 to 0.9 or below. Functional analysis of the 171 proteins was performed using the Cytoscape plug-in, BiNGO [63] to assess the over-representation of associated Gene Ontology molecular function terms. Gene Ontology Slim definitions for *Saccharomyces cerevisiae* [64] were used discarding the evidence codes IEA (inferred from electronic annotation), ISS (inferred from sequence structural similarity) and NAS (non-traceable author statement). A hypergeometric test with false discovery rate correction [65] was used to select and visualize the significantly enriched GO functions at a level p<0.001, using the GO-s of the entire network as reference set. Colors represent functional categories: red, nucleus-related; blue, transport-related; green, signaling-related; yellow denotes other functions. The size of the circles represents the number of proteins found in the category.

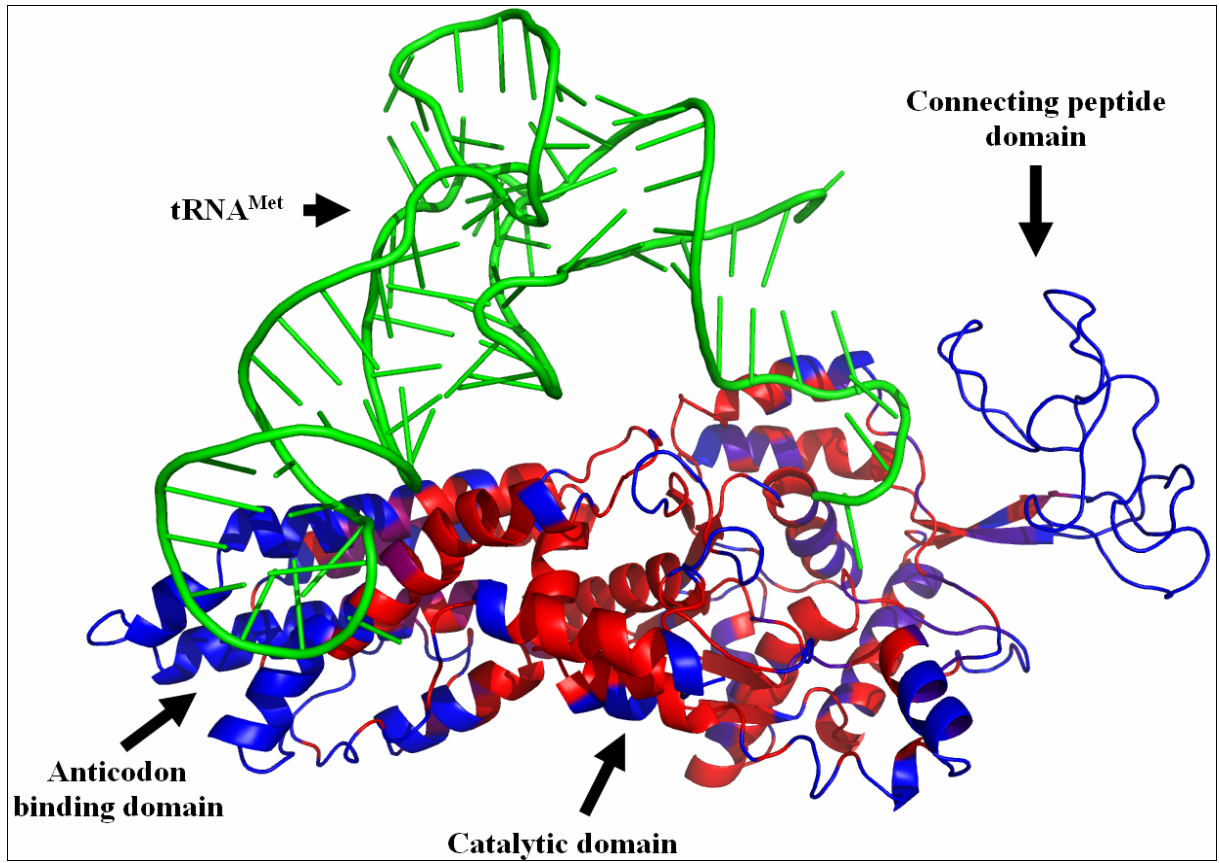

**Figure 2. Game centralities of *E. coli* methionyl-tRNA synthetase amino acids.** The protein structure network of *E. coli* methionyl-tRNA-synthetase was constructed, prisoner's dilemma game was simulated, and game centrality measures were calculated as described in Methods. Game centralities were overlaid to the 3D image of the protein and tRNA made by the PyMOL program package [66]. tRNA$^{Met}$ is shown in green, the most influential amino acids spreading defection are marked red (these amino acids have the largest game centrality, GC values) and the least influential amino acids are blue (having the smallest GC values).

# Supporting Information (Text S1)

# Nodes having a major influence to break cooperation define a novel centrality measure: game centrality

**Gabor I. Simko[1,2], Peter Csermely[1,*]**

**1** Department of Medical Chemistry, Semmelweis University, Budapest, Hungary,
**2** Vanderbilt University, Nashville, TN USA

*Corresponding author, E-mail: csermely.peter@med.semmelweis-univ.hu

## Summary

In this Supporting Information (Text S1) we give a detailed description of the NetworGame spatial social dilemma game simulation program package. Besides the pseudocode description of the NetworGame algorithm the Supporting Information also contains a supplementary figure, 2 supplementary tables as well as 11 references.

**The computer programs of the NetworGame package with a User Guide can be downloaded from here: www.linkgroup.hu/NetworGame.php.**

## Table of Contents

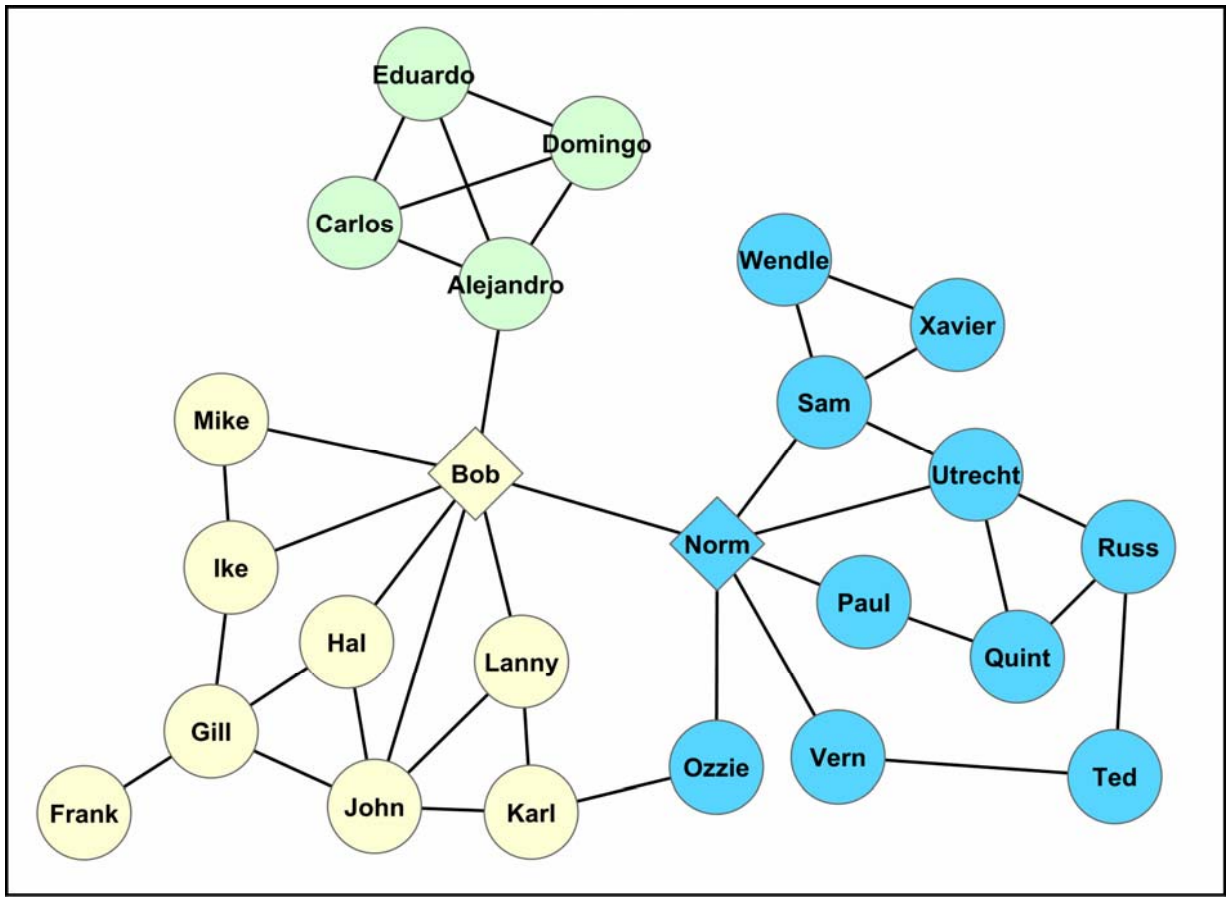

**Supplementary Figure S1.** Michael's strike network [1]. The three worker groups of a former forest product manufacturing factory containing younger, English-speaking (yellow); older, English-speaking (blue); or younger, Spanish-speaking workers (green) were marked. Sam and Wendle (top right) were the union leaders, who failed to break the strike, while Bob and Norm (center, marked with diamonds) were the pair of workers, who successfully broke the strike.

**Supplementary Tables**

**Table S1. List of consensus party hubs**

**Consensus party hub ORFs[a]**

| | |
|---|---|
| YAR002W | YHR077C |
| YAR003W | YHR089C |
| YBL004W | YHR166C |
| YBL007C | YHR200W |
| YBL038W | YIL115C |
| YBL050W | YJR045C |
| YBL084C | YJR065C |
| YBL099W | YJR068W |
| YBR010W | YJR121W |
| YBR084W | YKL018W |
| YBR087W | YKL022C |
| YBR118W | YKL068W |
| YBR245C | YKL085W |
| YDL029W | YKL129C |
| YDL065C | YLR127C |
| YDL134C | YLR212C |
| YDL208W | YMR080C |
| YDL213C | YMR109W |
| YDR103W | YMR116C |
| YDR118W | YNL016W |
| YDR244W | YNL094W |
| YDR264C | YNL102W |
| YDR395W | YNL138W |
| YER157W | YNL172W |
| YFR002W | YNL290W |
| YFR036W | YOL094C |
| YGL004C | YOR157C |
| YGL153W | YOR249C |
| YGL200C | YOR250C |
| YHL030W | YOR270C |
| YHR016C | YPL213W |
| | YPR088C |

[a]The open reading frame names of 63 consensus yeast party hubs were determined and listed as in [2] comparing the party hubs of the high fidelity yeast protein-protein interaction network [3] with those published in other 5 publications [4-8], and listing only those as 'consensus party hubs', which were never classified as a date hub.

**Table S2**. List of consensus date hubs

**Consensus date hub ORFs**[a]

| | | | |
|---|---|---|---|
| YAL005C | YER095W | YKL081W | YNL093W |
| YBL016W | YER110C | YKL095W | YNL127W |
| YBL023C | YER148W | YKL104C | YNL135C |
| YBL093C | YER155C | YKL166C | YNL243W |
| YBL105C | YER165W | YKL203C | YNL263C |
| YBL106C | YFL017W-A | YKR001C | YNL271C |
| YBR011C | YFR021W | YKR026C | YNL298W |
| YBR089C-A | YFR028C | YKR068C | YOL086C |
| YBR114W | YFR034C | YLL021W | YOL090W |
| YBR119W | YGL003C | YLL026W | YOL108C |
| YBR126C | YGL092W | YLL039C | YOL123W |
| YBR135W | YGL116W | YLR044C | YOL133W |
| YBR160W | YGL198W | YLR096W | YOL135C |
| YBR175W | YGL207W | YLR180W | YOR039W |
| YBR254C | YGR009C | YLR229C | YOR089C |
| YBR274W | YGR040W | YLR310C | YOR106W |
| YBR279W | YGR086C | YLR319C | YOR212W |
| YCR009C | YGR104C | YLR337C | YOR244W |
| YDL047W | YGR134W | YLR342W | YOR304W |
| YDL101C | YGR218W | YLR423C | YOR308C |
| YDL126C | YGR274C | YLR452C | YPL004C |
| YDL160C | YHR061C | YML007W | YPL031C |
| YDL188C | YHR099W | YML010W | YPL082C |
| YDR142C | YHR152W | YML064C | YPL129W |
| YDR155C | YIL038C | YML109W | YPL153C |
| YDR170C | YIL046W | YMR001C | YPL161C |
| YDR172W | YIL094C | YMR012W | YPL181W |
| YDR192C | YJL081C | YMR043W | YPL248C |
| YDR216W | YJL095W | YMR054W | YPL256C |
| YDR238C | YJL138C | YMR125W | YPR054W |
| YDR240C | YJL141C | YMR139W | YPR072W |
| YDR309C | YJL164C | YMR199W | YPR086W |
| YDR473C | YJL187C | YMR201C | YPR107C |
| YDR523C | YJL194W | YMR213W | YPR119W |
| YEL009C | YJR066W | YMR273C | YPR182W |
| YER081W | YJR090C | YMR304W | |
| | YJR091C | YNL006W | |

[a]The open reading frame names of 145 consensus yeast date hubs were determined and listed as in [2] comparing the date hubs of the high fidelity yeast protein-protein interaction network [3] with those published in other 5 publications [4-8], and listing only those as 'consensus date hubs', which were never classified as a party hub.

**Description of the NetworGame algorithm**

The 2.0 version of the NetworGame program is an updated version of the NetworGame 1.0 version published in a preliminary conference report [9]. NetworGame 2.0 is available in our web-site (www.linkgroup.hu/NetworGame.php). The 2.0 version utilizes our experiences gained with the 1.0 version. The NetworGame 2.0 program package is a cross-platform, generic tool to simulate repeated spatial games. This simulation program includes i.) options for pay-off matrices of any symmetric normal form games (with 2 strategies); ii.) several well-known, replicator-type strategy update rules, as well as the option for additional, user-defined strategy update rules in a 'plugin'-type format; iii.) synchronous, and semi-synchronous updating [10]; iv.) and the option for the inclusion of any real world networks in a Pajek format [11].

Here we provide the pseudocode for the algorithm, which describes the flow of the program and the effects of the configuration parameters. A User Guide of version 2.0 can be downloaded from here: www.linkgroup.hu/NetworGame.php.

**Configurator**
 – testNode and testEdge are configuration parameters
 – printSteps, printStepsStdDev and printLast are configuration parameters
 – Nodes and Edges represent the network, where Edges is a set of pairs (src,dst)

```
initialize payoff matrix
if (testNode specified) then
  for i in Nodes do
    initialize strategies
    Si = testNode
    run simulations
    print statistics
  end
else if (testEdge specified) then
  for (src,dst) in Edges do
    initialize strategies
    Ssrc = testEdge
    Sdst = testEdge
    run simulations
    print statistics
  end
else
  initialize strategies
  run simulations
  if (printSteps) then print step-wise average cooperation levels
  if (printStepsStdDev) then print step-wise standard deviances
  if (printLast) then print average cooperation for each node at last step
end
```

**Run simulations**
- M is a set of simulations
- L is number of steps
- n is the size of M
- memUsage and elapsedTime are internal variables representing the current memory usage of the system and the elapsed time since the start of the simulations
- maxError is a parameter controlling the statistical accuracy
- numberofsimulations, numberofsteps, mem and time are configuration parameters

```
if (numberofsimulations specified) then
  n = numberofsimulations, i.e. M has size of numberofsimulations
else
  n = 100, i.e. M has size of 100 initially, but it can grow
end

if (numberofsteps specified) then
  L = numberofsteps is the number of steps in each simulation
else
  L = 101 (and it can grow)
end
```

  – *run simulations until we reach the specified resource limits, or get below the desired statistical error*
```
while (memUsage < mem and elapsedTime < time) do
  for m in M do
    simulate m up to steps L
  end
  A_i = average cooperation at step i for each m in M
  currMean = average of A_x, where x=L-50 to L
  prevMean = average of A_x, where x=L-150 to L-100
  stddev   = standard deviation of A_x, where x=L-50 to L
  meanerror = sqrt(stddev / n*(n-1))
  if (numberofsteps unspecified and abs(prevmean-currmean) > maxError) then
    L = L + 1
  else if (numberofsimulation unspecified and meanerror > maxError/2) then
    add a simulation to M (inherently n = n + 1)
  else
    finish simulations
  end
end
```
  – *calculate statistics for printing*
```
for i = 1,2...,L do
  calculate average cooperation at step i over all m in M
  calculate standard deviation at step i over all m in M
end

for n in Nodes do
  calculate average cooperation at last step for n over all m in M
end
```

## Simulate m up to steps L

- $S_i$ is the current strategy of node i
- $P_i$ is the current payoff of node i
- useWeights, x0 and x1 are weight parameters controlling the effect of edge weights
- Neighbors(i) is the set of neighbors for node i
- Payoff[i,j] is the payoff matrix value when strategy j plays against strategy j
- payoffSchema is a configuration parameter

```
for n = 1,2,...,L do
```
  – *simulating current round and calculating payoffs*
```
    for i in Nodes do
      P_i = 0
      counter = 0
```
  – *the probability of a game is dependent on the weight parameters and edge weight $W_{i,j}$*
```
      for j in Neighbors(i) do
        if (not useWeights or random(0,1) <= (W_{i,j}-x0)/(x1-x0)) then
          P_i = P_i + Payoff[S_i,S_j]
          counter = counter + 1
        end
        if (payoffSchema = degree or payoffSchema = averaging) then
          P_i = P_i / counter (if counter > 0)
      end
```
  – *updating strategies (strategyUpdateRule can be implemented as a plugin, it may have memory, or might be one of the built-in rules: best takes over or proportional update*
```
    for k in Nodes do
      S_k = strategyUpdateRule(...)
    end
end
```